\documentclass[12pt, twocolumn]{article}%
\usepackage{amssymb}
\usepackage{amsmath,latexsym}
\usepackage{graphicx}
\usepackage{amsmath}
\usepackage{amsfonts}%
\begin{document}

\title{{\Huge Classical electron model with non static conformal symmetry}}
\author{I. Radinschi$^{\text{1}}$, F. Rahaman$^{\text{2}}$, M. Kalam$^{\text{3}}$ and
K. Chakraborty$^{\text{4}}$\\$^{\text{1}}$Department of Physics, "Gh. Asachi" Technical University,\\Iasi,
 700050, Romania\\$^{\text{2}}$Dept. of Mathematics, Jadavpur University,\\Kolkata-700 032,
  India\\$^{\text{3}}$Dept. of Physics, Aliah
University, Sector - V , Salt Lake,  Kolkata - 700091,
India\\$^{\text{4}}$ Dept. of Physics, Govt. Training College,
Hooghly - 712103, West Bengal, India}
\date{}
\maketitle

\begin{abstract}
Lorentz proposed a classical model of electron in which electron
was assumed to have only 'electromagnetic mass'.  We modeled
electron as charged anisotropic perfect fluid sphere admitting
non static conformal symmetry. It is noticed  that the pressure
and density fail to be regular at the origin but effective
gravitational mass is regular everywhere and vanishes at the
limit $r\rightarrow0$ i.e. it does not have to tolerate the
problem of singularity. Further, we have matched interior metric
with exterior (Reissner-Nordstr\"{o}m) metric and determine the
values of the parameters $k$ and $r_{0}$ ( occurring in the
solutions ) in functions of mass, charge and radius of the
spherically symmetric charged objects i.e. electron.

\end{abstract}

\footnotetext{

$^{\text{1}}${radinschi@yahoo.com }
\par
$^{\text{2}}$farook\_rahaman@yahoo.com
\par
$^{\text{3}}$ mehedikalam@yahoo.co.in
\par
$^{\text{4}}$ kchakraborty28@yahoo.com}
\par

\section{lntroduction}

In recent years, Herrera and Varela [1] discussed the electron model assuming
the equation of state $p_{radial}+\rho=0$ with ad hoc anisotropy. Ray et al
[2] studied electron model admitting one parameter group of conformal motion.
We also notice that a new electromagnetic mass model admitting Chaplygin gas
equation of state with particular specialization was developed [3]. In this
study, we extend both the work in [1] and [2].

Valuable studies performed in the last two decades point out that a very
interesting topic is the charged imperfect fluid spheres with a space-time
geometry that admits a conformal symmetry, both in the static and the
generalized nonstatic cases [4], [5], [6], [7], [8], [9] (and references cited
therein). \newline These classes of solutions are used in relativistic
astrophysics for developing the star models. Herrera and his collaborators
[5], [6], [7] searched for exact solutions of the field equations for static
spheres considering that the static and spherically symmetric fluid geometry
also presents a conformal symmetry. Ponce de Leon [8] gave regular static
solutions with anisotropic pressure in the approach of a conformally flat
sphere. Maartens and Maharaj [4] looked more deeply and obtained regular
solutions of the Einstein-Maxwell equations for charged imperfect fluids with
conformal symmetry.

We investigate a new electron model as a charged anisotropic
perfect fluid sphere admitting non static conformal symmetry.
 The paper is organized
as follows: in Section 2 we give the basic equations which are
the Einstein-Maxwell field equations combined with the
electromagnetic tensor field with anisotropic fluid. In Section 3
the solutions and some particular cases are given together with
the graphical representation of the results. In Section 4 we
match interior metric with the exterior (Reissner-Nordstr\"{o}m)
metric and find out $k$ and $r_{0}$ parameters in terms of mass,
charge and radius of the spherically symmetric charged  objects
i.e. electron. Section 5 is devoted to a discussion of the
results.

\section{Basic Equations for the New Electron mass Model with Non Static
Conformal Symmetry}

The mass of the electron originates from its electromagnetic
field itself. The electron is modeled as a charged fluid sphere
obeying Einstein-Maxwell equations. To perform our study, we
consider the Einstein-Maxwell field equations combined with the
electromagnetic tensor field with anisotropic fluid, which
together with an appropriate equation of state (EOS),
$p_{radial}+\rho=0$ yield the basic equations used for developing
a new electron model with non static conformal symmetry.

The most general energy momentum tensor compatible with spherically symmetry
is
\begin{equation}
T_{\nu}^{\mu}=(\rho+p_{r})u^{\mu}u_{\nu}+p_{r}g_{\nu}^{\mu}+(p_{t}-p_{r }%
)\eta^{\mu}\eta_{\nu}, \label{Eq3}%
\end{equation}
with

$u^{\mu}u_{\mu}=-\eta^{\mu}\eta_{\mu}=1,$

where $\rho$ is the matter density, and $p_{r}$ and $p_{t}$ are the radial
pressure and the transverse pressure of the fluid, respectively.

The static spherically symmetric space-time is given by the line-element (in
geometrized units with $G=1=c$)
\begin{equation}
ds^{2}=-e^{\nu(r)}dt^{2}+e^{\lambda(r)}dr^{2}+r^{2}(d\theta^{2}+sin^{2}\theta
d\phi^{2}), \label{Eq3}%
\end{equation}

where the functions of radial coordinate $r$,$\ \nu(r)$ and $\lambda(r)$ are
the metric potentials. \newline\newline\newline\newline For this metric, the
Einstein Maxwell field equations are
\begin{equation}
e^{-\lambda}\left[  \frac{\lambda^{\prime}}{r}-\frac{1}{r^{2}}\right]
+\frac{1}{r^{2}}=8\pi\rho+E^{2},
\end{equation}%
\begin{equation}
e^{-\lambda}\left[  \frac{1}{r^{2}}+\frac{\nu^{\prime}}{r}\right]  -\frac
{1}{r^{2}}=8\pi p_{r}-E^{2},
\end{equation}%
\[
\frac{1}{2}e^{-\lambda}\left[  \frac{1}{2}(\nu^{\prime})^{2}+\nu^{\prime
\prime}-\frac{1}{2}\lambda^{\prime}\nu^{\prime}+\frac{1}{r}({\nu^{\prime
}-\lambda^{\prime}})\right]  =8\pi p_{t}+E^{2}.
\]
\begin{equation}
\end{equation}

where $p_{i}$, $\rho$ and $E(r)$ represent fluid pressures ( radial and
transverse ), matter-energy density and electric field, respectively.

The electric field is expressed by
\begin{equation}
(r^{2}E)^{\prime}=4\pi r^{2}\sigma e^{\frac{\lambda}{2}}.
\end{equation}

where $\sigma$ represents the charge density of the spherical
distribution.

The equation (6) gives the following form for the electric field
\begin{equation}
E(r)=\frac{1}{r^{2}}\int_{0}^{r}4\pi r^{2}\sigma e^{\frac{\lambda}{2}}%
dr=\frac{q(r)}{r^{2}},
\end{equation}
with $q(r)$ the total charge of the sphere under consideration.

Like most of the researchers who studied the electron model, we assume the
equation of state
\begin{equation}
p_{r}=-{\rho}. \label{Eq3}%
\end{equation}
The assumption (8) is consistent with the 'Causality Condition' $|\frac
{dp}{d\rho}|\leq1$ [10]. It is very common for cosmologists to use the matter
distribution obeying this type of equation of state with the equation of state
parameter $\omega=-1$ (known as false vacuum or $\rho$ - vacuum) to explain
the acceleration phase of the Universe [11].

\section{Solutions}

Now, we consider  electron as a charged fluid sphere under
conformal motion through non static Conformal Killing Vector as
[4]
\begin{equation}
L_{\xi}g_{ij}=g_{ij;k}\xi^{k}+g_{kj}\xi_{;i}^{k}+g_{ik}\xi_{;j}^{k}=\psi
g_{ij}, \label{eqckv}%
\end{equation}

where $L$ represents the Lie derivative operator, $\xi$ is the
four vector along which the derivative is taken, $\psi$ is the
conformal factor, and $g_{ij}$ are the metric potentials [9].
\newline\newline\newline For a vanishing $\psi$ the equation
above yields the Killing vector, the case $\psi=const.$
corresponds to the homothetic vector, and for $\psi=\psi(x,t)$ we
obtain conformal vectors. In this way Conformal Killing Vector
allows a more general study of the space-time geometry.

The proposed charged fluid ( electromagnetic mass ) space time is mapped
conformally onto itself along the direction $\xi$.

Here, one takes $\xi$ non static but $\psi$ is static as
\begin{equation}
\xi=\alpha(t,r)\partial_{t}+\beta(t,r)\partial_{r}, \label{eqK1}%
\end{equation}%
\begin{equation}
\psi=\psi(r) \label{eqK1}%
\end{equation}
The above equations give the following set of expressions [4]
\begin{equation}
\alpha=A+\frac{1}{2}kt, \label{eqK1}%
\end{equation}%
\begin{equation}
\beta=\frac{1}{2}Bre^{-\frac{\lambda}{2}}, \label{eqK1}%
\end{equation}%
\begin{equation}
\psi=Be^{-\frac{\lambda}{2}}, \label{eqK1}%
\end{equation}%
\begin{equation}
e^{\nu}=C^{2}r^{2}exp\left[  -2kB^{-1}\int\frac{e^{\frac{\lambda}{2}}}%
{r}dr\right]  , \label{eqK1}%
\end{equation}
where $C$, $k$, $A$, $B$ are constants. According to [4] one can set, $A=0$
and $B=1$ by re-scaling.

The equation of state (8) implies
\begin{equation}
\nu=-\lambda. \label{eqK1}%
\end{equation}

Equations (15) and (16) yield
\begin{equation}
-\lambda^{\prime}=\frac{2}{r}-2k\frac{e^{\frac{\lambda}{2}}}{r}. \label{eqK1}%
\end{equation}
Solving this equation, one gets,
\begin{equation}
e^{\frac{\lambda}{2}}=\frac{1}{\left(  k-\frac{r}{r_{0}}\right)  },
\label{eqK2}%
\end{equation}
where $r_{0}$ is an integration constant.

We discuss the model by assuming the following assumption.
\begin{equation}
\sigma e^{\frac{\lambda}{2}}=\sigma_{0}r^{s}, \label{eqK2}%
\end{equation}
where  $\sigma_{0}$, $\lambda$ and $s$ represent the charge
density at the center of the system, metric potential and a
constant, respectively. Usually, the term $\sigma
e^{\frac{\lambda}{2}}$ inside the integral sign in equation (7)
is known as volume charge density. One can interpret the
assumption (19) as the volume charge density being polynomial
function of r [12].

Thus finally, we obtain the following set of solutions for different
parameters
\begin{equation}
e^{\lambda}=e^{-\nu}=\frac{1}{\left(  k-\frac{r}{r_{0}}\right)  ^{2}},
\label{eqK2}%
\end{equation}%
\begin{equation}
\psi=\left(  k-\frac{r}{r_{0}}\right)  , \label{eqK1}%
\end{equation}%
\begin{equation}
E(r)=\left[  \frac{4\pi\sigma_{0}}{(s+3)}\right]  r^{s+1},
\end{equation}%
\begin{equation}
q(r)=\left[  \frac{4\pi\sigma_{0}}{(s+3)}\right]  r^{s+3},
\end{equation}
$8\pi\rho=\left(  k-\frac{r}{r_{0}}\right)  ^{2}\left[  \frac{2}{r(kr_{0}%
-r)}-\frac{1}{r^{2}}\right]  \newline+\frac{1}{r^{2}}-\left[  \frac{16\pi
^{2}\sigma_{0}^{2}}{(s+3)^{2}}\right]  r^{2s+2},$
\begin{equation}
\end{equation}
$8\pi p_{r}=\left(  k-\frac{r}{r_{0}}\right)  ^{2}\left[  \frac{1}{r^{2}%
}-\frac{2}{r(kr_{0}-r)}\right]  \newline-\frac{1}{r^{2}}+\left[  \frac
{16\pi^{2}\sigma_{0}^{2}}{(s+3)^{2}}\right]  r^{2s+2},$
\begin{equation}
\end{equation}%
\begin{equation}
8\pi p_{t}=\frac{1}{r_{0}^{2}}\left[  3-\frac{2(kr_{0}-r)}{r}\right]  -\left[
\frac{16\pi^{2}\sigma_{0}^{2}}{(s+3)^{2}}\right]  r^{2s+2}.
\end{equation}

It is worthwhile to calculate effective gravitational mass which
is due to the contribution of the energy density $\rho$ of the
matter  and the electric energy density $\frac{E^2}{8\pi}$ and
can be expressed as

$M=\int_{0}^{r}4\pi r^{2}\left[  \rho+\frac{E^{2}}{8\pi}\right]
dr\newline=\frac{k}{r_{0}}r^{2}+(1-k^{2})\frac{r}{2}-\left(  1+\frac{1}%
{2}k^{2}\right)  \frac{r^{3}}{3r_{0}^{2}}.$
\begin{equation}
\end{equation}
It is noted that the metric potentials $\nu(r)$ $\&$ $\lambda(r)$
, $\psi$ ,  electric field $E(r)$ , electric charge $q(r)$ ,
matter energy density and pressures are depended on several
unknown parameters such as  on $\sigma_{0}$, $s$, $k$ and
$r_{0}$. So, to get the exact values of these parameters one will
have to match the solutions with  Reissner-Nordstr\"{o}m metric
and will be discussed later.

We have $p_{r} =- \rho$ which implies that a negative value of
the matter energy density $\rho$ determines a positive value of
the radial pressure $p_{r}$.\\ Here, the NEC energy condition
$\rho+p_{r}\geqslant0$ is satisfied (  for equality with zero ).
The violation of NEC implies the breakdown of causality in
general relativity and the violation of the second law of
thermodynamics [13]. The condition $\rho+p_{t}>0$ is satisfied
for $\left( k-\frac{r}{r_{0}}\right) ^{2}\left[
\frac{2}{r(kr_{0}-r)}-\frac {1}{r^{2}}\right]  + \frac{1}{r^{2}}
\newline+ \frac{1}{r_{0}^{2}}\left[
3-\frac{2(kr_{0}-r)}{r}\right]  >\left[  \frac{32\pi^{2}\sigma_{0}^{2}%
}{(s+3)^{2}}\right]  r^{2s+2} $

This means there exists a limiting value of the radial coordinate
for which $\rho+p_{t}>0$.

One can note that apparently  there is no singularity at $r=0$
for the metric coefficients, $\sigma$, $E$ if $s>-1$.
\\
\\
\\
\\
\\
But the  Kretschmann scalar

$K=R_{\mu\nu\alpha\beta}R^{\mu\nu
\alpha\beta}=\\
\frac{4}{r^{4}r_{0}^{4}} [10r^{2}k^{2}r_{0}^{2}-12r^{3}kr_{0}+6r^{4}+r_{0}%
^{4}-2r_{0}^{4}k^{2}+4r_{0}^{3}kr-2r_{0}^{2}r^{2}+k^{4}r_{0}^{4}-4k^{3}%
r_{0}^{3}r]$

 indicates that there is a point of  divergence at  $r=0$.
 The pressure and density also fail to be
regular at the origin but the effective gravitational mass is
always positive and will vanish as $r\rightarrow 0$ i.e. it does
not have to tolerate the problem of singularity.

\section{ Matching with Reissner Nordstr\"{o}m metric}

To match interior metric with the exterior (Reissner-Nordstr\"{o}m) metric, we
impose only the continuity of $g_{tt}$, $g_{rr}$ and $\frac{\partial g_{tt}%
}{\partial r}$, across a surface, S at $r= a $%

\begin{equation}
1-\frac{2m}{a}+\frac{Q^{2}}{a^{2}}=\left(  k-\frac{a}{r_{0}}\right)  ^{2},
\end{equation}

\begin{equation}
\frac{m}{a^{2}}-\frac{Q^{2}}{a^{3}}=-\frac{1}{r_{0}}\left(  k-\frac{a}{r_{0}%
}\right)  .
\end{equation}
From, the above three equations, one could find the values of the
unknown $k$ and $r_{0}$ in terms of mass, charge and radius of
electron as
\begin{equation}
r_{0}=-\frac{\left[  1-\frac{2m}{a}+\frac{Q^{2}}{a^{2}}\right]  ^{\frac{1}{2}%
}}{\left(  -\frac{m}{a^{2}}+\frac{Q^{2}}{a^{3}}\right)  },
\end{equation}%
\begin{equation}
k=\frac{\left[  1-\frac{3m}{a}+\frac{2Q^{2}}{a^{2}}\right]  }{\left[
1-\frac{2m}{a}+\frac{Q^{2}}{a^{2}}\right]  ^{\frac{1}{2}}}.
\end{equation}
According Herrera et al [1], we assume the following values of
the electron as:  radius $a\sim10^{-16}$ cm , the inertial mass
$m\sim10^{-56}$cm and charge $Q\sim10^{-34}$cm in relativistic
units. So the matching conditions imply
\begin{equation}
r_{0}=-10^{20}%
\end{equation}%
\begin{equation}
k=1
\end{equation}

In Fig.1, Fig.2, Fig.3, Fig.4, Fig.5 and Fig.6, we plot electric
charge $q(r)$, electric field $E(r)$, matter energy density
$\rho$, radial pressure $p_{r}$, effective gravitational mass of
the electron $M$ and the transverse pressure of the fluid $p_{t}$
against the $r$ parameter, assuming the above values of the
parameters $r_{0}$ and k.\newline

\section{Discussions}
   A
new electron model as a charged anisotropic perfect fluid sphere
admitting non static conformal symmetry is investigated and in
this view we performed our calculations with the Einstein-Maxwell
field equations combined with the electromagnetic tensor field of
an anisotropic fluid and with an appropriate equation of state
(EOS), $p_{radial}+\rho=0$. For describing the behavior of our
model, we assume that $\sigma
e^{\frac{\lambda}{2}}=\sigma_{0}r^{s}$ and obtain the expressions
for the metric potentials $\nu(r)$ and $\lambda(r)$, $\psi$, the
electric field $E(r)$, the electric charge $q(r)$, the matter
energy density $\rho$, the radial pressure $p_{r}$, the
transverse pressure of the fluid $p_{t}$, and the effective
gravitational mass of the electron $M$. It is noticed  that the
pressure and density fail to be regular at the origin but
effective gravitational mass is regular everywhere and vanishes
at the limit $r\rightarrow0$ i.e. it does not have to tolerate the
problem of singularity.  We extend our study by matching interior
metric with the exterior (Reissner-Nordstr\"{o}m) metric, and
also establish the values of the parameters $k$ and $r_{0}$  in
terms of the mass, charge and radius of the  electron. . Using
the values obtained for $r_{0}$ and k, one finds the charge of
the fluid is nearly equal to $q\sim10^{-34}$ cm in relativistic
units  which is equivalent to the charge of electron (
experimentally obtained [1] ).  The matching of interior metric
with exterior (Reissner-Nordstr\"{o}m) metric,
$1-\frac{2m}{a}+\frac{Q^{2}}{a^{2}}=1-\frac{2M}{a}$, where M is
the effective gravitational mass, implies $M=m-\frac{Q^{2}}{2a}$.
If we use the above values of inertial mass, charge and radius of
the electron, one gets the value of the effective gravitational
mass within the fluid sphere as $\sim-10^{-52}$cm. Using the
above obtained values of $r_{0}$ and k, one finds that the
effective gravitational mass of the charged fluid is nearly equal
to $\sim-10^{-52}$ (see Fig. 5). \newline In the present work, we
have used non static conformal symmetry technique to search
charged fluid of radius $\sim10^{-16}$. It may be interesting to
extrapolate the present investigation to the astrophysical
bodies, especially quarks or strange stars.

\section{Acknowledgements}

FR is thankful to PURSE for financial support. We are also
thankful to Dr. S. Ray, Dr. A Bhattacharya and Prof. S Chakraborty
for several Illuminating discussions. We are very grateful   to
an anonymous referee for his/her insightful comments that have
led to significant improvements, particularly on the
interpretational aspects.

\pagebreak

\begin{figure}[ptbhptbhptbh]
\centering  \includegraphics[scale=.3]{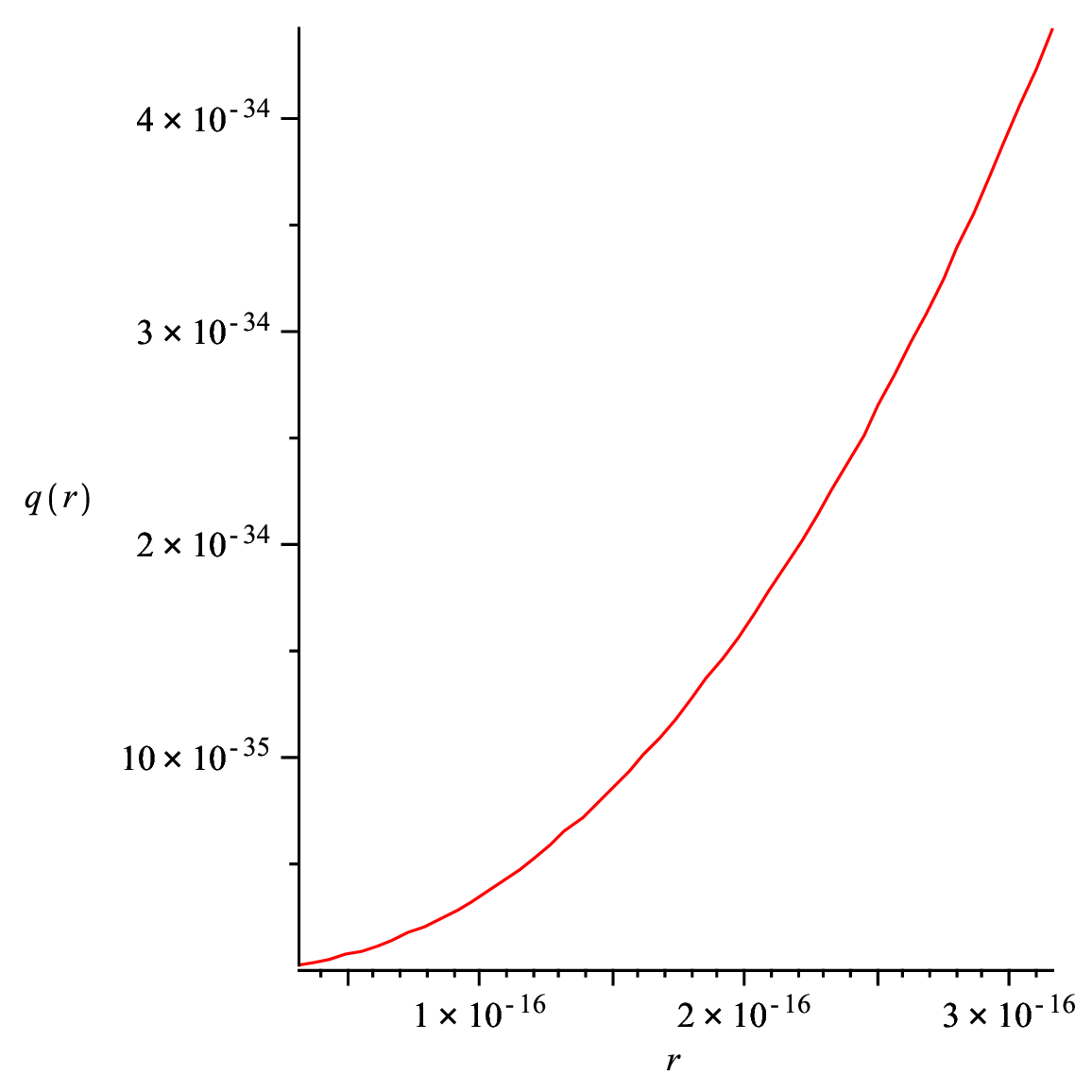} \caption{The diagram of the
Electric charge with respect to radial coordinate 'r' for s = -.8 and
$\sigma_{0} =1$. }%
\label{fig:wormhole}%
\end{figure}\begin{figure}[ptbhptbhptbhptbh]
\centering  \includegraphics[scale=.3]{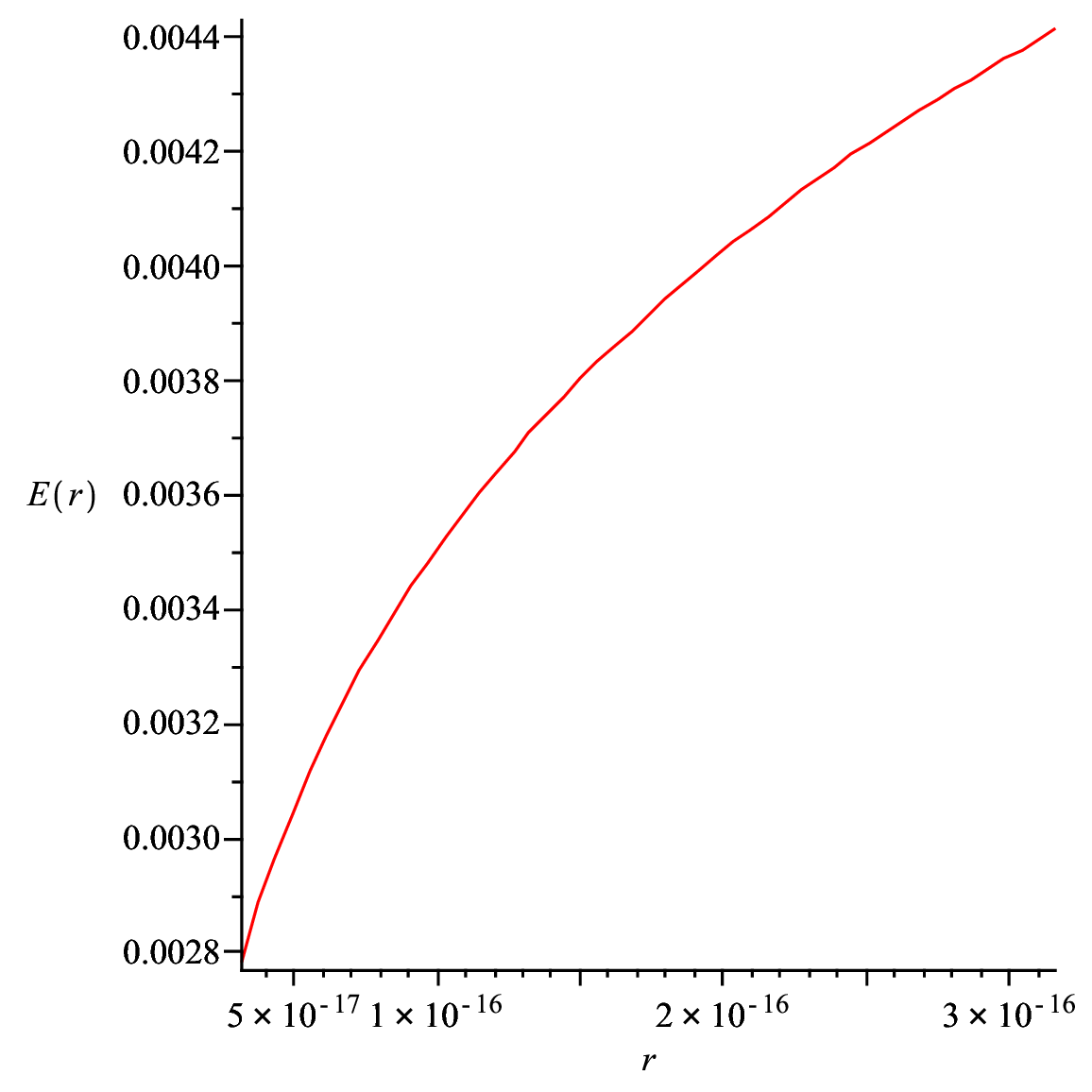} \caption{The diagram of the
Electric field strength with respect to radial coordinate 'r' for s = -.8 and
$\sigma_{0} =1$. }%
\label{fig:wormhole}%
\end{figure}

\begin{figure}[ptbh]
\centering  \includegraphics[scale=.4]{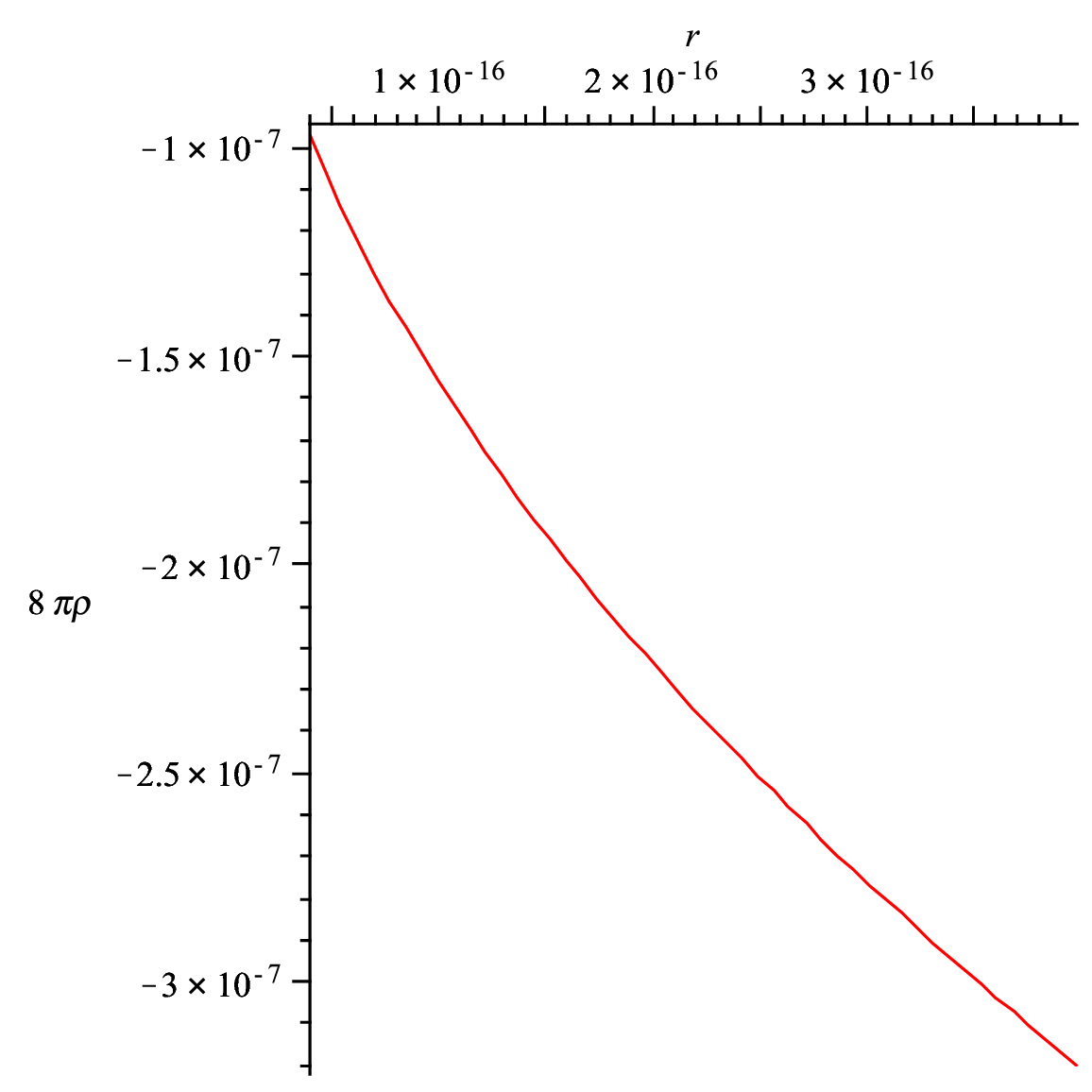} \caption{The diagram of the
energy density with respect to radial coordinate 'r' for s = -.8 and
$\sigma_{0} =1$. }%
\label{fig:wormhole}%
\end{figure}\begin{figure}[ptbhptbh]
\centering  \includegraphics[scale=.4]{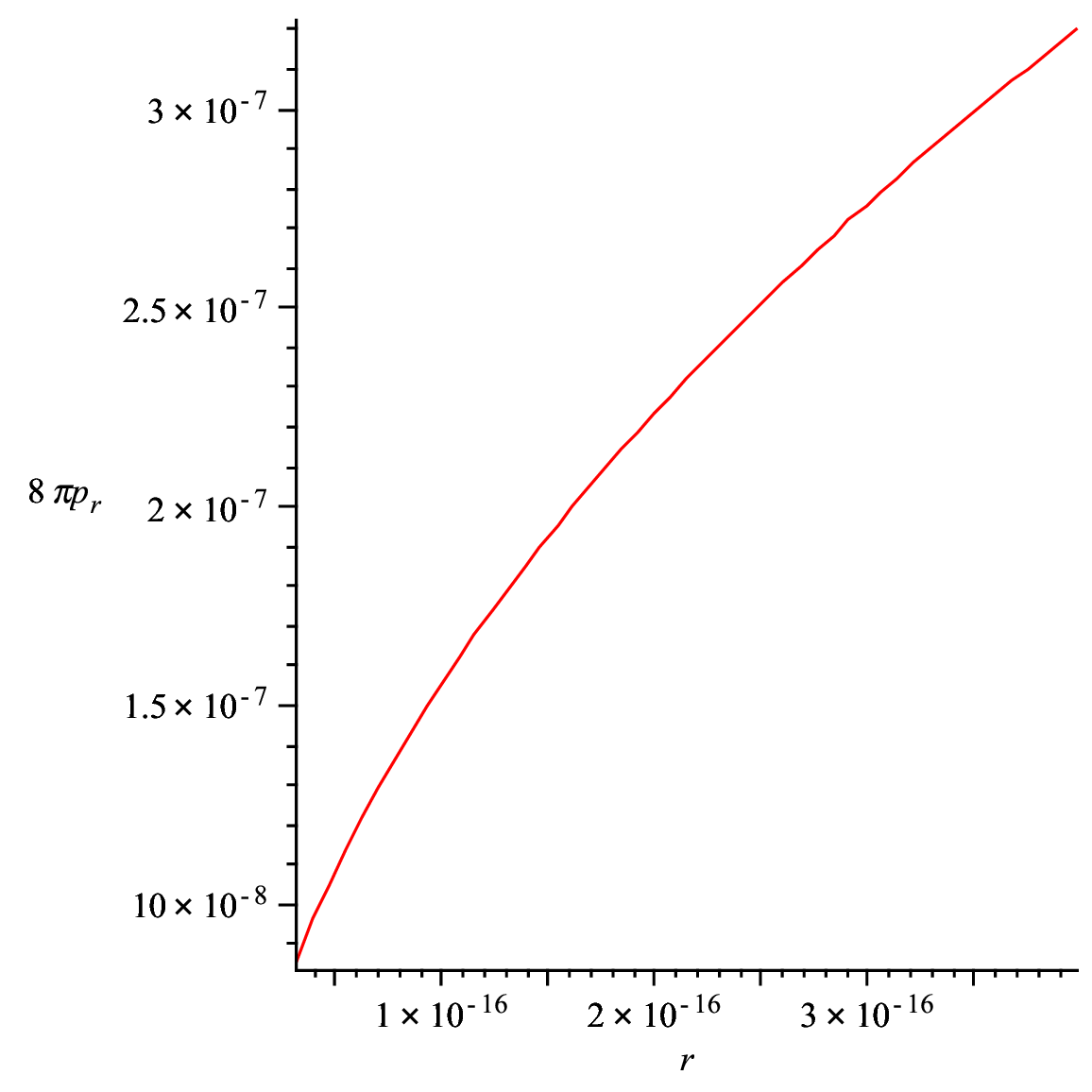} \caption{The diagram of the
radial pressure with respect to radial coordinate 'r' for s = -.8 and
$\sigma_{0} =1$. }%
\label{fig:wormhole}%
\end{figure}\begin{figure}[ptbhptbhptbh]
\centering  \includegraphics[scale=.4]{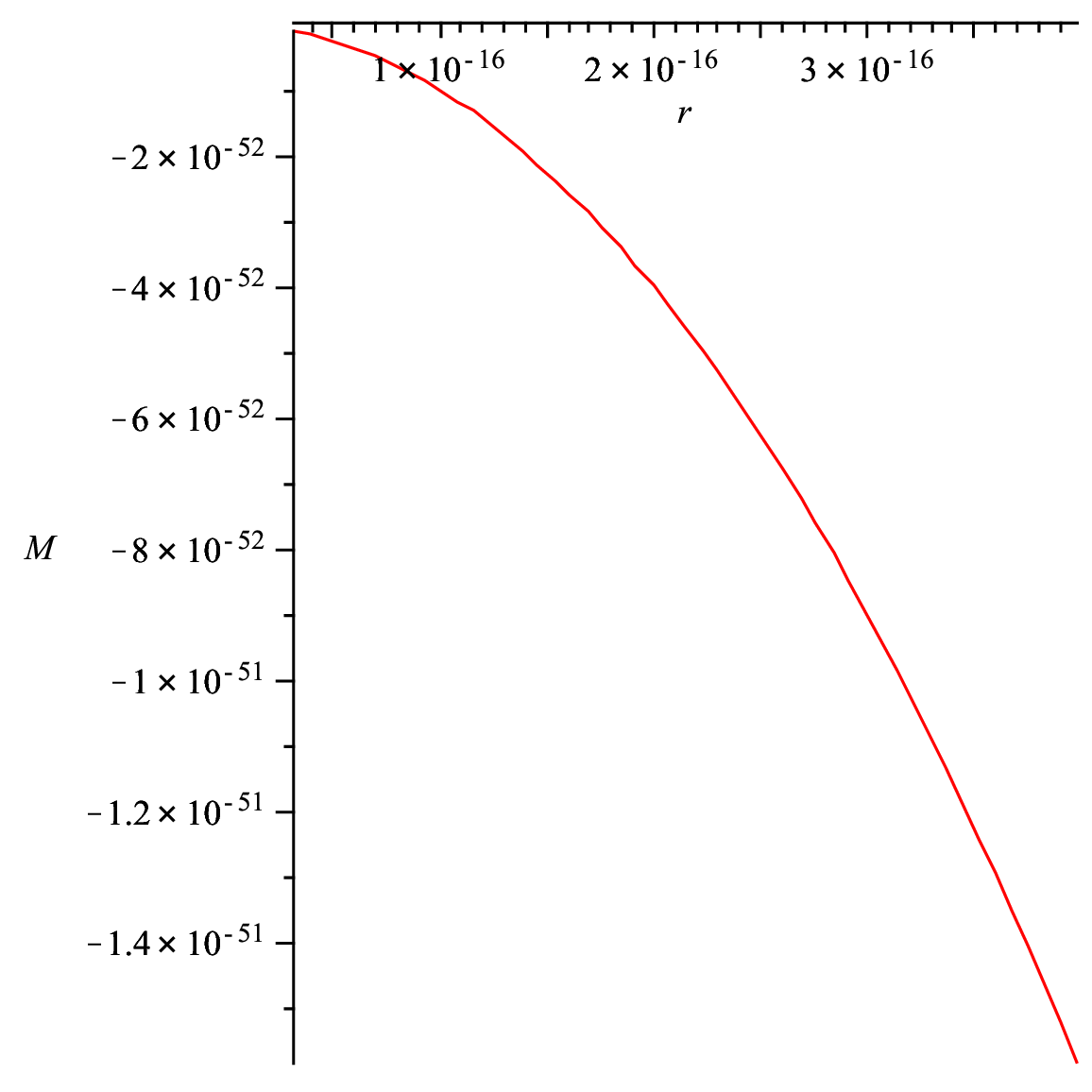} \caption{The diagram of the
effective gravitational mass function with respect to radial coordinate 'r'. }%
\label{fig:wormhole}%
\end{figure}\begin{figure}[ptbhptbhptbhptbh]
\centering  \includegraphics[scale=.4]{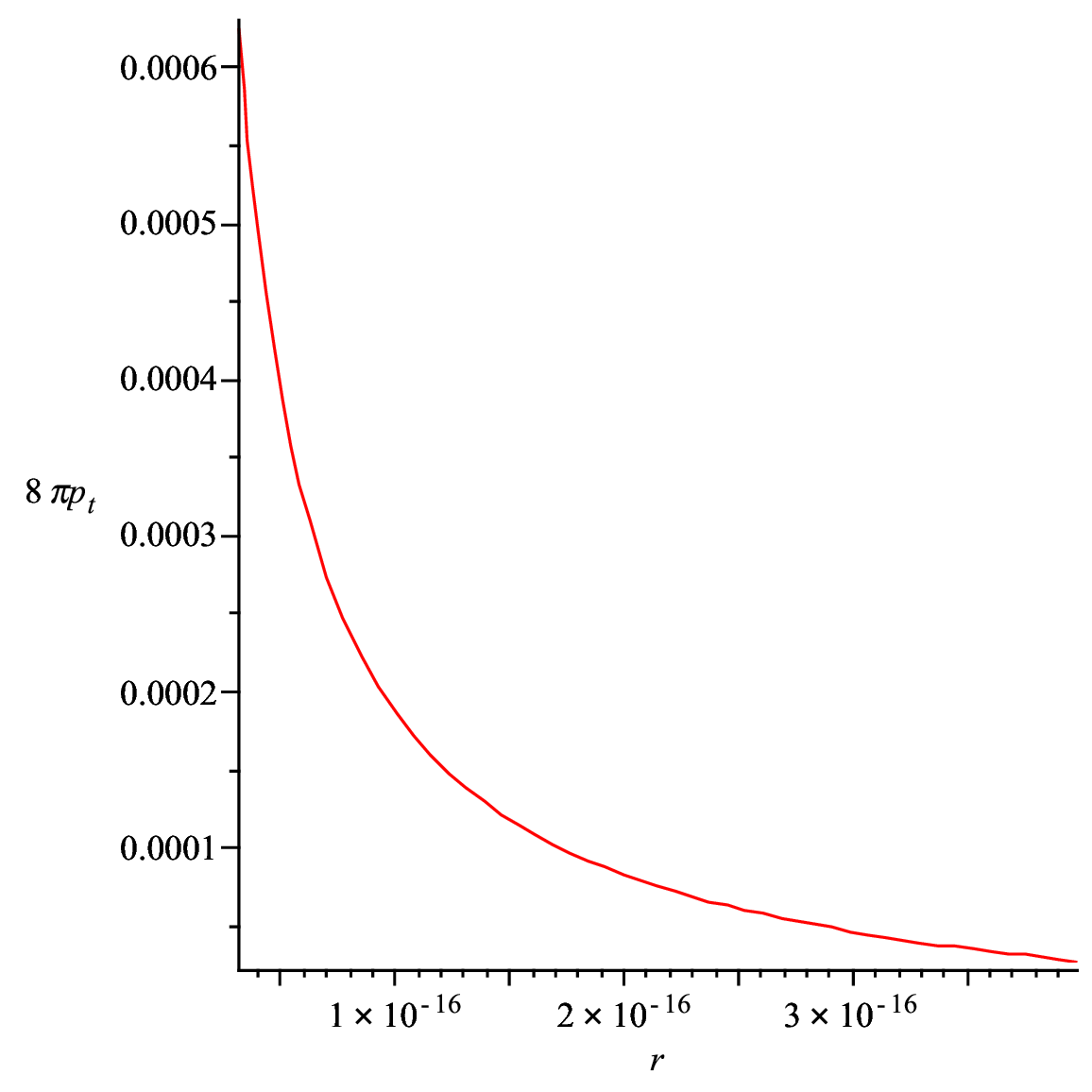} \caption{The diagram of the
transverse pressure with respect to radial coordinate 'r' for s = -.8 and
$\sigma_{0} =1$. }%
\label{fig:wormhole}%
\end{figure}


\begin{thebibliography}{99}                                                                                               %


\bibitem {1}S. L. Herrera and V. Varela, Phys. Lett. A 189 (1994)
11; \\
W B Bonnor and F I Cooperstock {\it Phys. Lett. A} {\bf 139}
(1989)442
\bibitem {1}S. Ray, A. A. Usmani, F. Rahaman, M. Kalam and K. Chakraborty,
Ind.J.Phys. 82, 1191 (2008), arXiv: 0806.3568[gr-qc]

\bibitem {3}I. Radinschi, F. Rahaman, M. Kalam and K. Chakraborty, arXiv: 0811.0068[hep-th]

\bibitem {4}R. Maartens and M. S. Maharaj, J. Math. Phys. 31 (1990) 151.

\bibitem {5}L. Herrera, J. Jimenez, L. Leal, J. Ponce de Leon, M. Esculpi and
V. Galina, J. Math. Phys. 25 (1984)3274.

\bibitem {6}L. Herrera and J. Ponce de Leon, J. Math. Phys. 26(1985)2302 .

\bibitem {7}R. Maartens, D. P. Mason and M. Tsamparlis, J. Math. Phys. 27 (1986)2987.

\bibitem {8}J. Ponce de Leon, J. Math. Phys. 28 (1987) 1114.

\bibitem {9}F. Rahaman et al, \newline arXiv:0808.2927[astro-ph]

\bibitem {10}N. R J Adler, J.Math.Phys. 15 (1974) 727.

\bibitem {10}N. V Sahni et al, arxiv: astro-ph/9904398

\bibitem {10}F. Rahaman et al,\newline Int.J.Theor.Phys.48:471-475,2009, arXiv:0807.4596[gr-qc]

\bibitem {10}N. Arkani-Hamed, S. Dubovsky, A. Nicolis, E. Trincherini and G.
Villadoro, JHEP 0705 (2007) 55.
\end{thebibliography}
\end{document}